\documentstyle[12pt]{article}
\begin{document}

\newcommand{\newc}{\newcommand}
 
\newc{\be}{\begin{equation}}
\newc{\ee}{\end{equation}}
\newc{\ba}{\begin{eqnarray}}
\newc{\ea}{\end{eqnarray}}
\newc{\bea}{\begin{eqnarray}}
\newc{\eea}{\end{eqnarray}}
\newc{\D}{\partial}
\newc{\ie}{{\it i.e.} }
\newc{\eg}{{\it e.g.} }
\newc{\etc}{{\it etc.} }
\newc{\etal}{{\it et al.} } 
\newc{\ra}{\rightarrow}
\newc{\lra}{\leftrightarrow}
\newc{\no}{Nielsen-Olesen }
\newc{\lsim}{\buildrel{<}\over{\sim}}
\newc{\gsim}{\buildrel{>}\over{\sim}}
 
\begin{titlepage}
\begin{center}
December 1997\hfill CRETE-97/20 \\ 
\hfill CWRU-P22-97 \\

\vskip 0.5cm
 
{\large \bf PHASE TRANSITIONS IN THE CORE OF GLOBAL EMBEDDED DEFECTS}

\vskip .1in
{\large Minos Axenides}\footnote{E-mail address:
 axenides@gr3801.nrcps.ariadne-t.gr },\\[.05in]
 {\em Institute of Nuclear Physics,\\ N.C.R.P.S. Demokritos \\
153 10, Athens, Greece 
}
\vskip .1in
{\large Leandros Perivolaropoulos}\footnote{E-mail address:
leandros@physics.uch.gr},\\[.05in]
{\em Department of Physics\\
University of Crete\\
71003 Heraklion, Greece
}
and
\vskip 0.1in
{\large Mark Trodden}\footnote{E-mail address:
trodden@theory1.phys.cwru.edu},\\[.05in]
{\em Particle Astrophysics Theory Group\\ Department of Physics\\ Case
Western Reserve University\\ Cleveland, OH 44106-7079 }
\end{center}
\vskip .1in
\begin{abstract}
\noindent 
We demonstrate the existence of global monopole and vortex configurations 
whose core exhibits 
a phase structure. We determine the critical values of parameters for which 
the transition from the symmetric to the non-symmetric phase occurs and 
discuss 
the novel dynamics implied by the non-symmetric cores for defect interactions.
We model phase transitions in the core of global embedded topological defects 
by identifying the relevant parameters with the vaccuum expectation value of 
a dynamical scalar field. 
Finally, we argue that superheavy defects that undergo a core phase transition 
in the 
very early universe provide a novel realization for topological inflation. 
\end{abstract} 
\end{titlepage}

\section{Introduction}
\noindent
The existence of topological defects\cite{vs94,v85} with a non-trivial
core phase structure has recently been demonstrated for embedded global 
domain walls and vortices
\cite{ap97} (see also \cite{sz83} for earlier studies).
The analogous phenomenon has also been found to occur in experiments 
performed in superfluid
$^3He$ (see e.g. Ref. \cite{mv95}). In that case, a phase 
transition in the core of 
superfluid $^3He-B$ vortices
was observed to correspond to different spin-orbital states of their core.
In the case of the field theory, the partial breaking of global symmetries 
\cite{ap97} deforms the vacuum manifold in such a way as to reduce the 
dimensionality of the non-contractible configurations (points,circles,spheres 
etc.) that are otherwise admitted, giving rise to "embedded" topological 
defects (domain walls-vortices).
In these defects a core with a nontrivial 
symmetry phase structure generically appears as a consequence of the 
embedding. 
This structure is associated with either the zero value of a scalar field in 
the core 
(symmetric phase) or a nonzero value (broken phase) 
for different regions
of the enlarged parameter space that the vacuum manifold deformation 
affords.
Such embedded defects may be viewed\cite{ap97} as interpolating between
``texture'' type defects\cite{s61,t89,bt95}, where the field boundary
conditions are uniform, and symmetric core defects\cite{t74,no73},
where the topological charge is due to the field variation at the
boundaries. Superconducting strings \cite{w85} are a special case
of this type of defect, in which the core acquires a nonzero vacuum
expectation value (VEV) for a particular range of parameters. The
existence of a non-symmetric core implies a large variety
of new properties and phenomena for such topological defects. Some of
these effects were in fact discussed in Ref. \cite{ap97}, where it was 
shown,
for example, that in the range of parameters for which a non-symmetric
core is favored, an initially spherical domain wall bubble is
unstable towards planar collapse. That study focused on the
stability and properties of domain walls and vortices with
a  non-symmetric core.

In the present work we complete the list of such defects by presenting a 
model for an ``embedded'' global monopole\cite{bv89}. Moreover we carefully 
study the implications of non-symmetric cores for defect interactions 
such as those between vortices and those between monopoles.

In addition, we investigate the stability of new types of core structures
with non-trivial winding in the vortex core describing vortices within 
vortices. However, as we show, these structures are unstable. In the next 
section, we construct an $ O(4)$ model that admits global monopoles with a 
core exhibiting both symmetric and nonsymmetric false vacuum phases for
appropriate values of the free parameters, and identify their range
of values for which this core structure is stable. We also
analyze the vortex-vortex and monopole-monopole interactions for
defects with non-symmetric cores. We show that the non-trivial core
structure induces an additional interaction potential which can be
attractive or repulsive depending on the relative orientation of the
scalar field value in the defect cores.

An interesting implication of our results is that the transition from a 
symmetric to the non-symmetric false vacuum
in the defect core can be viewed as a genuine
{\it phase transition} if the parameters of our models are rendered new 
scalar external degrees of freedom
with their own dynamics. Symmetric or non-symmetric cores correspond to the 
zero or nonzero value of
a scalar field condensate developing in the defect core. In a topological 
inflation scenerio, an inflating
string core \cite{topinf} enters a re-heating period if a non-zero VEV
develops in its core, rapidly decreasing the vacuum energy there.
This possibility is studied in section 4.

\section{Vortex Core Dynamics}
Before discussing the monopole case, we briefly review and extend the
case of the vortex with non-trivial core structure first discussed in
Ref. \cite{ap97}. Consider the following Lagrangian density describing
the symmetry breaking $SU(2) \rightarrow U(1) \rightarrow I$, where the
first breaking is explicit and the second is spontaneous:
\be
{\cal{L}} = {1 \over 2} \D_\mu {\Phi^\dagger}
\D^\mu \Phi + {M^2 \over 2} {\Phi^\dagger}\Phi + {m^2 \over 2}
{\Phi^\dagger}
\tau_3 \Phi - {h\over 4} ({\Phi^\dagger}\Phi)^2 \ .
\ee
Here, $\Phi = (\Phi_1, \Phi_2)$ is a complex scalar doublet and
$\tau_3$ is the $2 \times 2$ Pauli matrix. Using the rescaling
 
\begin{eqnarray}
\Phi &\rightarrow &{M\over \sqrt{h}} \Phi \\
x & \rightarrow & {1\over M} x \\ m & \rightarrow & \alpha M \ ,
\label{rescale1}
\end{eqnarray}
\noindent
the potential reduces to
\be
V(\Phi) = -{M^4 \over {2 h}} ((\Phi^\dagger \Phi) + \alpha^2
(|\Phi_1|^2 -|\Phi_2|^2)- {1\over 2} (\Phi^\dagger \Phi)^2) \ .
\ee
The ansatz

\be
\Phi =
\left( \begin{array}{c}
\Phi_1 \\
\Phi_2
\end{array} \right)=
\left( \begin{array}{c}
f(\rho) e^{i m \theta} \\ g(\rho)
\end{array}
\right) \ ,
\ee
\noindent
with boundary conditions
\bea
\lim_{\rho \rightarrow 0} f(\rho)
&=& 0, \hspace{3cm} \lim_{\rho \rightarrow 0} { g^\prime} (\rho) = 0
\\
\lim_{\rho \rightarrow \infty} f(\rho) &=& (\alpha^2 +1)^{1/2},
\hspace{1cm} \lim_{\rho \rightarrow \infty} g (\rho) = 0 \ ,
\eea
\noindent
leads to the field equations
\bea
f'' + {f' \over \rho} - {m^2 \over \rho^2} f +(1 + \alpha^2)f - (f^2 +
g^2) f &=& 0
\\
g'' + {g' \over \rho} + (1-\alpha^2) g - (f^2 +g^2)g &=& 0 \ .
\eea
\noindent
It is straightforward \cite{ap97} to use a relaxation algorithm to
numerically solve the system (9,10) with the boundary conditions (7,8).
For
$\alpha > \alpha_{cr} \simeq 0.37$ the solution relaxes to a form that
is symmetric at the defect core ($g(0)=0$), while for $\alpha <
\alpha_{cr}$ a non-zero VEV at the core is energetically
favored. As a check of the numerical results, we have also verified the 
stability of the symmetric ($g(\rho)=0$) solution by solving the 
linearized eigenvalue problem of
equation (10) around $g(\rho)=0$ and showing that there are no
negative modes for $\alpha > \alpha_{cr}$.

\noindent
A more interesting core structure is obtained through the ansatz
\noindent
\be
\Phi =
\left( \begin{array}{c}
\Phi_1 \\
\Phi_2
\end{array} \right)=
\left( \begin{array}{c}
f(\rho) e^{i m \theta} \\ g(\rho) e^{i n \phi} \ .
\end{array}
\right)
\ee
\noindent
The resulting configuration corresponds to the formation of an
additional
$\Phi_2$ vortex localized within the core of the $\Phi_1$ vortex. This
`vortex within vortex' configuration can be metastable in the
parameter region where a non-zero VEV of the $\Phi_2$ component is
favored in the core of the $\Phi_1$ vortex. Clearly, since the true
vacuum outside the vortex has $\Phi_2 = 0$, the winding can only
persist within the core of the $\Phi_1$ vortex. The ansatz (11) leads
to the same field equation (9) for $f(\rho)$,  while the equation for
$g(\rho)$ becomes
\noindent
\be
g'' + {g' \over \rho} - {n^2 \over \rho^2} g^2 + (1-\alpha^2) g - (f^2
+g^2)g = 0 \ ,
\ee
\noindent
with boundary conditions
\noindent
\bea
\lim_{\rho \rightarrow 0} f(\rho)
&=& 0, \hspace{3cm} \lim_{\rho \rightarrow 0} { g} (\rho) = 0
\\
\lim_{\rho \rightarrow \infty} f(\rho) &=& (\alpha^2 +1)^{1/2},
\hspace{1cm} \lim_{\rho \rightarrow \infty} g (\rho) = 0 \ .
\eea
\noindent
Using a relaxation method\cite{numrec} again, we find that for all
$\alpha
> 0$, the system (9), (12) relaxes to a solution with $g(\rho) =
0$. Thus, a stable non-trivial winding does not develop within the
vortex core. We have verified this result by examining the field
equation 
solution
\noindent
\be
\Phi =
\left( \begin{array}{c}
\Phi_1 \\
\Phi_2
\end{array} \right)=
\left( \begin{array}{c}
f(\rho) e^{i m \theta} \\ 0
\end{array}
\right)
\ee
for unstable modes towards the configuration (11).  This amounts to
solving the
linearized eigenvalue problem
\noindent
\be
-g'' - {g' \over \rho} + {n^2 \over \rho^2} g^2 - (1-\alpha^2) g + f^2
g = \omega^2 g \ ,
\ee
\noindent
where $f$ satisfies equation (9) with $g=0$, and looking for negative
eigenvalues $\omega^2$. Using a shooting method \cite{numrec}, with
initial conditions $g(0)=0$ and $g'(0) = 1$ (required for non-trivial
winding), we showed that, for nonzero winding, $n$, $g$ has no
instability 
for any $\alpha$. The stability of such {\it `vortices within
vortices'} however, is expected to improve by introducing a $U(1)$
gauge field. This effectively screens the repulsive centrifugal
barrier ${n^2 \over \rho^2} g^2$ which prohibits the development of a
negative mode and drives the system towards a `vortex within vortex'
configuration. A detailed general study of defects
within defects is currently in progress\cite{p97}.

The existence of non-trivial structures within the core of defects has
interesting implications for their interactions. For
example, consider a vortex produced by an explicit breaking of $O(3)$
symmetry to $O(2)$, with subsequent spontaneous breaking to $I$. The
corresponding vortex ansatz is of the form

\be
\Phi =
\left( \begin{array}{c}
\Phi_1 \\
\Phi_2 \\
\Phi_3
\end{array} \right)=
\left( \begin{array}{c}
f(\rho) \cos\phi \\
\pm f(\rho) \sin\phi \\
g(\rho)
\end{array}
\right) \ ,
\ee
\noindent
with boundary conditions
\bea
\lim_{\rho \rightarrow 0} f(\rho)
&=& 0, \hspace{3cm} \lim_{\rho \rightarrow 0} { g} (\rho) = \pm c
\\
\lim_{\rho \rightarrow \infty} f(\rho) &=& const ,
\hspace{2cm} \lim_{\rho \rightarrow \infty} g (\rho) = 0 \ .
\eea
\noindent 
Here, the $+$ ($-$) (in (17)) corresponds to a vortex (antivortex), and
$c$ is determined dynamically and depends on the parameters of the
Lagrangian. Consider now a field configuration corresponding to a pair
of vortices at a large distance $l$ from each other, and let $\rho_1$,
$\rho_2$ be the distances of any point from the centers of the two
vortices respectively. The corresponding field configuration is of the
form

\be
\Phi =
\left( \begin{array}{c}
\Phi_1 \\
\Phi_2 \\
\Phi_3
\end{array} \right)=
\left( \begin{array}{c}
f(\rho_1)f(\rho_2) \cos(\phi_1 \pm \phi_2)\\ f(\rho_1)f(\rho_2)
\sin(\phi_1 \pm \phi_2)\\ g(\rho_1,\rho_2)
\end{array}
\right) \ ,
\ee
\noindent
with $\vert g(0,l)\vert = \vert g(l,0) \vert = c$ and the $+$, ($-$)
corresponds to a vortex-vortex (vortex-antivortex) pair. Consider
first the case of a vortex-antivortex pair with $g_2 (0,l) = -g_2
(l,0) = \pm c$. Assuming a linear interpolation of $g_2$ between $g_2
(0,l)$ and $g_2 (l,0)$, and substituting the total energy of the sum of
two isolated vortices for the total energy of the interacting
configuration (see also Ref \cite{p92}) we find, for large separations
$l$, the interaction energy
\be
E_{int}^{V-{\bar V}} \simeq 2\pi \eta (\log (\eta l) - log(\eta L)) +
A{{2c^2}\over l} \ .
\ee
\noindent
Here, $\eta$ is the scale of symmetry breaking, $A$ is a positive
constant with dimensions of length squared and $L$ is a cutoff length
related to the size of the system. We have also assumed that the cores
remain undistorted during the interaction. The form of $E_{int}^{V-{\bar
V}}$
implies that the attractive vortex-anti-vortex force could be balanced
by the repulsive force originating from the gradient energy of the $g$
component at a critical distance

\be
l_{cr} = {{A c^2} \over {\eta \pi}} \ .
\ee
\noindent
This however cannot lead to a vortex-anti-vortex static bound state,
because, according to Derrick's theorem, no static, finite energy scalar
field configuration exists in two dimensions with non-zero potential
energy. This means that the above expression for the interaction
energy breaks down at distances larger than $l_{cr}$.

A similar interaction term can be obtained by a vortex-vortex pair
 with $g_2 (0,l) = g_2 (l,0) = \pm c$. In this case it is
 straightforward to show that $E_{int}^{V-V} \simeq -E_{int}^{V-{\bar
V}}$
 and there is an attractive gradient-reducing term balancing the
 standard repulsive vortex-vortex interaction at the same critical
 distance given by (22). We have shown however\cite{p97}, that despite 
 the fact that this configuration
 has infinite energy and Derrick's theorem cannot be applied, defect
 bound states do not exist in this case either.  The detailed
 investigation of these issues, with and without gauge fields is
 currently in progress \cite{p97} using numerical simulations of
 defect evolution.

\section{Monopole Core Dynamics}
The above study of the core structure and interactions of vortices can
easily be extended to monopoles. As a concrete example we consider a
model with $O(4)$ symmetry explicitly broken to $O(3)$, which is
subsequently broken spontaneously to O(2). This is described by the
Lagrangian

\be
{\cal{L}} = {1 \over 2} \D_\mu {\Phi^\dagger}
\D^\mu \Phi + {M^2 \over 2} {\Phi^\dagger}\Phi + 
{m^2 \over 2} (|\Phi_1|^2 + |\Phi_2|^2 +|\Phi_3|^2) 
- {m^2 \over 2} |\Phi_4|^2
- {h\over 4} ({\Phi^\dagger}\Phi)^2 \ ,
\ee
\noindent  
with the monopole ansatz

\be
\Phi =
\left( \begin{array}{c}
\Phi_1 \\
\Phi_2 \\
\Phi_3 \\
\Phi_4 
\end{array} \right)=
\left( \begin{array}{c}
f(r) \cos\theta \sin\phi \\ f(r) \cos\theta \cos\phi \\ f(r)
\sin\theta \\ g(r)
\end{array}
\right) \ .
\ee
\noindent
After a rescaling similar to the one of equations (2,3,4), we obtain
the field equations for $f(r)$ and $g(r)$ as
\bea
f'' + {2 f' \over r} - {{2 f} \over r^2} +(1 + \alpha^2) - (f^2 + g^2)
f &=& 0
\\
g'' + {2 g' \over r} + (1-\alpha^2) g - (f^2 +g^2)g &=& 0 \ .
\eea

\begin{figure}
\begin{center}
\unitlength1cm
\begin{picture}(6,4)
\put(-6.5,7.7){\includegraphics{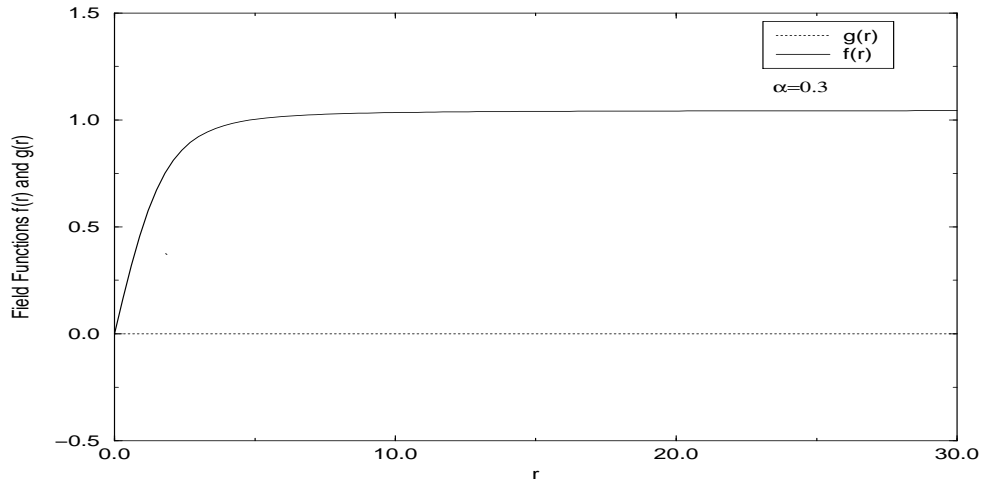}}
\end{picture}
\end{center}
\caption{Field configuration for a {\it symmetric-core}
global monopole with $\alpha = 0.3$.  }
\end{figure}

Using the same methods as for vortices, we find that the
development of a non-zero VEV for $g(r)$ at the monopole core is
energetically favored for
\be
\alpha < \alpha_{cr} \simeq 0.27
\ee
\noindent 
 As in the case of vortices this result was verified by two methods:
\begin{itemize}
\item
Using a relaxation method to solve the system (25,26) numerically with
boundary conditions
\bea
\lim_{r \rightarrow 0} f(r)
&=& 0, \hspace{3cm} \lim_{\rho \rightarrow 0} { g^\prime} (r) = 0
\\
\lim_{r \rightarrow \infty} f(r) &=& (\alpha^2 +1)^{1/2},
\hspace{1cm} \lim_{r \rightarrow \infty} g (r) = 0 \ .
\eea
The relaxed solution had $g(0) = 0$ for $\alpha >\alpha_{cr}$ (Fig. 1)
and $g(0) \neq 0$ for $\alpha <\alpha_{cr}$ (Fig. 2).
\item
Solving equation (25) numerically for $g=0$ and substituting in the
solution to (26), linearized in $g$. The resulting Schroedinger type
equation was then solved numerically with initial conditions $g'(0) =
0$ and $g(0) = 1$. For $\alpha >\alpha_{cr}\simeq 0.27$ it was shown
to have no negative eigenvalues. We found negative eigenvalues for
$\alpha <\alpha_{cr}$ thus showing the instability of the $g(r) = 0$
solution towards the solution obtained by the relaxation method ($g(0)
= c \neq 0$).
\end{itemize}

\begin{figure}
\begin{center}
\unitlength1cm
\begin{picture}(6,4)
\put(-6.5,7.7){\includegraphics{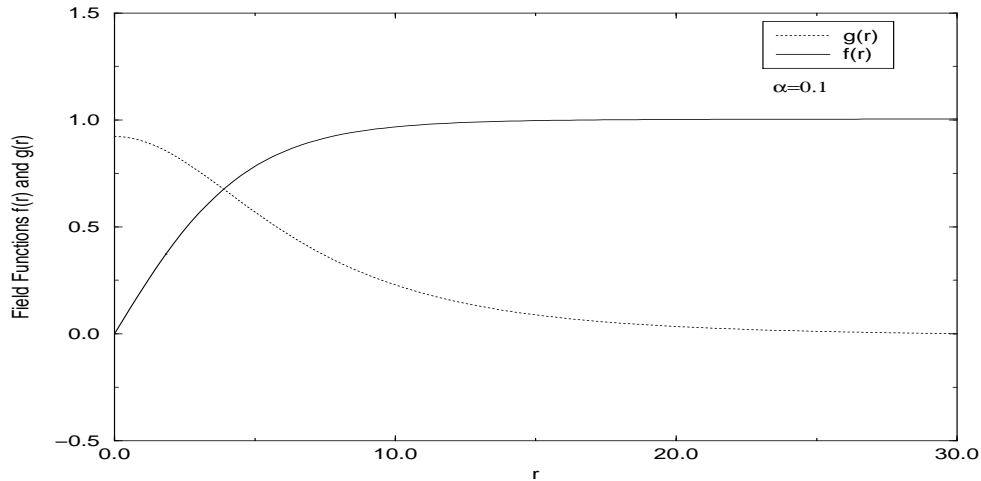}}
\end{picture}
\end{center}
\caption{Field configuration for a {\it non-symmetric-core}
global monopole with $\alpha = 0.1$.  }
\end{figure}

The monopole-antimonopole interactions can also be studied in a
similar way to vortex interactions. Using similar notation 
we find that the interacting configuration ansatz for a
monopole pair along the z axis is of the form

\be
\Phi =
\left( \begin{array}{c}
\Phi_1 \\
\Phi_2 \\
\Phi_3 \\
\Phi_4 
\end{array} \right)=
\left( \begin{array}{c}
f(r_1)f(r_2) \cos(\theta_1 \pm \theta_2) \sin\phi \\ f(r_1)f(r_2)
\cos(\theta_1 \pm \theta_2) \cos\phi \\ f(r_1)f(r_2) \sin(\theta_1 \pm
\theta_2) \\ g(r_1,r_2)
\end{array}
\right) \ ,
\ee
\noindent 
where the $+$ ($-$) sign corresponds to a configuration that would be
purely repeling (attracting) if it were not for the non-zero VEV of
$g(r_1,r_2)$ in the monopole cores.  Notice that in this case both
configurations correspond to a monopole-antimonopole pair as the
defects have opposite topological charges for both signs \cite{p92}.
An attracting configuration with $g(0,l)=-g(l,0)=\pm c$ has
interaction energy (for a derivation in the case of symmetric core see
Ref. \cite{p92}), at large separations $l$,
\be
E_{int}^- \simeq \eta^2 ((- 4 \pi^2 l + {{4 \pi L}\over 3}) + A {{2
c^2} \over l} \ ,
\ee
\noindent 
with $A$ a constant of dimension length squared.
\noindent 
Similarly for a repelling configuration with $g(l,0)=g(0,l)=\pm c $
the interaction energy is\cite{p92}
\be
E_{int}^+ \simeq \eta^2 (( 8 \pi^2 l + {{8 \pi L}}) - A {{2 c^2}
\over l} \ .
\ee
\noindent 
The precise form of these results requires numerical simulations of
defect evolution which are currently in progress\cite{p97}.

\section{Dynamical Core Symmetry Breaking and Topological Inflation}

It has recently been realized that superheavy topological defects (those 
for which the symmetry breaking scale is approximately the Planck scale)
can
act as seeds for inflation \cite{topinf}. This behavior is known as 
{\it topological
inflation} and occurs because at Planck scales, the core of a defect
consists of trapped vacuum energy density over a horizon-sized region
and
thus the slow-roll conditions are satisfied. In these scenarios, it is
natural to ask how inflation ends, since for traditional defects, if the 
slow-roll conditions are satisfied initially, then they will be
satisfied for 
all time as long as the evolution is classical.

However, for the types of defects we consider here, there is an 
alternative method for topological inflation to end. This occurs
if the defect is initially formed with a symmetric core 
(and hence has trapped energy density there), but at some later time 
the fields evolve in such a way that a non-symmetric core becomes
favored. 
In such a situation toplogical inflation should begin and then terminate
as
a field evolves through some critical value and there ceases to be
vacuum
energy density trapped in the core.

As a simple example, consider a modest extension of the vortex model 
mentioned earlier. The conditions necessary for topological 
inflation to take place are that the core be in the false vacuum and that
the
symmetry breaking scale be of order the Planck scale. We assume the
latter condition is true for the remainder of this section.
Introduce a new singlet scalar field, $\chi$, with
potential $U(\chi)$ and a $\chi^2{\Phi^\dagger}\tau_3\Phi$ coupling to
the $\Phi_i$. Note that we could introduce further couplings between the 
fields, consistent with the symmetries, but choose not to for the sake
of simplicity. The Lagrangian density is

\newpage

\begin{eqnarray}
{\cal{L}} = {1 \over 2} \D_\mu {\Phi^\dagger}
\D^\mu \Phi + {1 \over 2}(\partial_{\mu}\chi)\partial^{\mu}\chi -
\left({\lambda \over 2}\chi^2 \right. & + & \left. {m^2 \over 2}\right) 
{\Phi^\dagger}\tau_3\Phi + {M^2 \over 2} {\Phi^\dagger} 
\Phi \nonumber \\
& - & {h\over 4} ({\Phi^\dagger}\Phi)^2 - U(\chi) \ ,
\end{eqnarray}
where, here, $\alpha \equiv (m/M) > \alpha_{cr}\simeq 0.37$. We write
the 
potential for $\chi$ as

\be
U(\chi) = {a \over 4}\chi^4 -{b \over 3}\chi^3 +{c \over 2}\chi^2 \ ,
\ee
with $a(T),b(T),c(T) >0$, with $T$ temperature, and perform the
rescaling 
in (\ref{rescale1}) along with

\begin{eqnarray}
\chi \rightarrow {M\over \sqrt{h}} \chi \ , \ \ \ \ &
\lambda \rightarrow \lambda h \ , \ \ \ \ &  
a \rightarrow ah \ , \nonumber \\
b \rightarrow bM\sqrt{h} \ , \ \ \ \ & 
c \rightarrow cM^2 \ , \ \ \ \ &
{\cal L} \rightarrow {M^4 \over h} {\cal L} \ .
\end{eqnarray}
This yields the re-scaled Lagrangian density as

\begin{eqnarray}
{\cal L} = {1 \over 2}\D_\mu {\Phi^\dagger}
\D^\mu \Phi + {1 \over 2}(\partial_{\mu}\chi)\partial^{\mu}\chi & + &
{1 \over 2}(\alpha^2 -\lambda \chi^2)(|\Phi_1|^2-|\Phi_2|^2) +
{1 \over 2} {\Phi^\dagger}
\Phi \nonumber \\
& - & {1 \over 4} ({\Phi^\dagger}\Phi)^2 
-{a \over 4}\chi^4 +{b \over 3}\chi^3 -{c \over 2}\chi^2 \ .
\end{eqnarray}

the equations of motion for the three fields $\Phi_1$, $\Phi_2$ and
$\chi$ are
\begin{eqnarray}
\D^\mu \D_\mu \Phi_1 - (\alpha^2 -\lambda\chi^2 +1)\Phi_1 
+({\Phi^\dagger}\Phi)\Phi_1 & = & 0 \\
\D^\mu \D_\mu \Phi_2 - (1-\alpha^2 +\lambda\chi^2)\Phi_2 
+({\Phi^\dagger}\Phi)\Phi_2 & = & 0 \\
\D^\mu \D_\mu \chi + [\lambda({\Phi^\dagger}\tau_3\Phi)+c]\chi +a\chi^3
-
b\chi^2 & = & 0 \ .
\end{eqnarray}

Now, let us focus on the dynamics in the core of the defect. This is the
important region for topological inflation. At high temperatures, we
expect the potential for $\chi$ to have a global minimum at the origin. 
However, at low temperatures, we can arrange the potential to be a
simple quartic with local minimum at $\langle\chi\rangle=0$ and, for
appropriate parameter choices, a global minimum at 

\be
\chi_* \equiv \langle\chi\rangle_{min} \equiv {b \over 2a}\left[1+
\left(1-{4ac \over b^2}\right)^{1/2}\right] \ .
\ee

Let us choose $a$, $b$ and $c$ at low temperatures so that 

\bea
\alpha^2 -\lambda\chi_*^2 & < & \alpha_{cr} \\
| c | & > & \lambda v_2^2 \ ,
\eea
where $v_2$ is the values of $\Phi_2$ in the non-symmetric core (note
that
$\langle\Phi_1\rangle=0$ in the core always). 

The evolution of this system in the early universe should be as follows.
At a phase transition at around the Planck temperature, superheavy 
topological defects are formed, with a symmetric core and the auxiliary
field, $\chi$, trapped in its local minimum at the origin. The core of
such
a defect satisfies the conditions for topological inflation and the
space-time inside the core begins to expand exponentially. As the temperature
drops, the potential for $\chi$ evolves and at some time, $\chi=0$ becomes a 
quasi-stable state and at a later time tunnels and rolls to its global 
minimum at $\chi=\chi_*$. At that point the core of the defect undergoes a 
first order phase transition from symmetric to non-symmetric phase. When this 
happens, the conditions for topological inflation cease to be satisfied and 
inflation ends rapidly. Notice that this scenario predicts the existence of 
non-zero cosmological constant
$\Lambda$ because the defect core never reaches the global minimum of the
$\Phi$ potential.

This is a new scenario for ending topological inflation and takes
advantage of the new core structures we have introduced.
Similarly we can model a second order phase transition in the core of our 
defect by choosing $ b=0$.
At high temperatures the core is in the symmetric phase $<\chi > = 0$ which 
is a global minimun as before.
At low enough temperatures $ T < T_{crit} $ with  the parameter c relaxing 
continuously to negative values
$ c(T) < 0$
a nonzero scalar condensate  $<\chi>_{min} = -\frac{c}{a}$ develops 
{\it continuously} in the core. Its
trapped vacuum energy relaxes in contrast to the previous scenario 
continuously to the nonzero value
determined by the $\Phi$ field. 

\section{Conclusions}
We have demonstrated the existence of global vortex and monopole 
configurations which arise from vacuum manifolds $S^2$ and $S^3$ which are 
deformed by partial breaking of their global symmetries to $S^1$ and $S^2$ 
respectively. Furthermore, these {\it embedded topological defects} possess
either symmetric or non-symmetric cores for appropriately specified values of 
their parameters.  We have also shown that, without gauge fields, it is not 
possible to support specific metastable topological defects confined in the 
false vacuum of other defects. The existence of such confined defects in the 
presence of gauge fields is an open issue under investigation. Further, we 
have used arguments based on energetics to show that non-trivial structures
in defect cores can lead to additional types of interaction
potentials. Finally, we have extended our models in such a way that the
transition from symmetric to non-symmetric phase in the core is
dynamical, with a scalar field condensate being the effective order 
parameter. In that case, if topological inflation occurs in the symmetric 
defect core, a phase transition in the core may cause the vacuum energy
there to be reduced. This dynamical ingredient for topological 
inflation provides a novel exit mechanism for the defect core from its 
otherwise {\it eternally inflating} state and thus is a new mechanism for
terminating inflation.

\section{ Acknowledgements}
We thank T. Tomaras, T. Vachaspati and G. Volovic for discussions.
The work of L.P. and M.A. was supported by the EEC grants
$CHRX-CT93-0340$ and $CHRX-CT94-0621$ and is the result of a network 
supported by the European Science
Foundation.  The European Science Foundation acts as catalyst for the
development of science by bringing together leading scientists and
funding agencies to debate, plan and implement pan-European
initiatives. The work of M.T. was supported by the 
U.S. Department of Energy, the National Science foundation,
and by funds provided by Case Western Reserve University.

\end{document}